\documentclass[showpacs, letterpaper, preprint, preprintnumbers,amsmath,amssymb]{revtex4}
\usepackage{amsmath}
\usepackage{amssymb}
\usepackage{color}
\usepackage{mathrsfs}
\usepackage{graphics}
\usepackage{graphicx}
\usepackage{dcolumn}
\usepackage{bm}

\begin{document}
%\tightenlines
\def\cl{\centerline}
\def\ni{\noindent}
\def\ul{\underline}
\def\in{\indent}
\def\b{$\bullet$~~}
\def\hs{\hspace*{0.25in}}
\newcommand{\msun}{~M_\odot}
\def\lsim{\mathrel{\rlap{\lower4pt\hbox{\hskip1pt$\sim$}}
	\raise1pt\hbox{$<$}}} %less than or approx. symbol
\def\gsim{\mathrel{\rlap{\lower4pt\hbox{\hskip1pt$\sim$}}
	\raise1pt\hbox{$>$}}} %greater than or approx. symbol

\begin{center}
{\Large {The Physics of Neutron Stars}}
\\
\vskip 10pt
{\large {J.M. Lattimer and M. Prakash }}
\vskip 10pt 
Department of Physics and Astronomy \\
State University of New York at Stony Brook \\
Stony Brook, NY 11794-3800, USA.
\end{center}
\vskip 12pt
\parindent=20pt

\begin{quote}
Neutron stars are some of the densest manifestations of massive objects in the
universe.  They are ideal astrophysical laboratories for testing theories
of dense matter physics and provide connections among nuclear physics,
particle physics and astrophysics.  Neutron stars may exhibit
conditions and phenomena not observed elsewhere, such as
hyperon-dominated matter, deconfined quark matter, superfluidity and
superconductivity with critical temperatures near ${10^{10}}$ kelvin,
opaqueness to neutrinos, and magnetic fields in excess of $10^{13}$
Gauss.  Here, we describe the formation, structure, internal
composition and evolution of neutron stars.  Observations that include
studies of binary pulsars, thermal emission from isolated neutron
stars, glitches from pulsars and quasi-periodic oscillations from
accreting neutron stars provide information about neutron star masses,
radii, temperatures, ages and internal compositions. 
\end{quote}

The term neutron star as generally used today refers to a star with a
mass $M$ on the order of 1.5 solar masses 
$(M_\odot)$, a radius $R$ of $\sim 12$ km, and
a central density $n_c$ as high as 5 to 10 times the nuclear
equilibrium density $n_0 \simeq 0.16~{\rm fm}^{-3}$ of neutrons and
protons found in laboratory nuclei. A neutron star is thus one
of the densest forms of matter in the observable
universe~\cite{baym75,baym79,heiselberg00}. Although neutrons dominate the
nucleonic component of neutron stars, some protons (and enough
electrons and muons to neutralize the matter) exist. At supra-nuclear
densities, exotica such as strangeness-bearing
baryons~\cite{glendenning85,glendenning97}, condensed mesons (pion or
kaon)~\cite{picond79,kaplan86a,kaplan86b}, or even deconfined
quarks~\cite{collins75} may appear. Fermions, whether in the form of
hadrons or deconfined quarks, are expected to also exhibit
superfluidity and/or superconductivity.

Neutron stars encompass ``normal'' stars, with hadronic matter
exteriors in which the surface pressure and baryon density vanish (the
interior may contain any or a combination of exotic particles
permitted by the physics of strong interactions), and ``strange quark
matter'' (SQM) stars~\cite{alcock88}. An SQM star could have either a
bare quark matter surface with vanishing pressure but a large,
supra-nuclear density, or a thin layer of normal matter supported by
Coulomb forces above the quark surface. The name SQM star originates
from the conjecture that quark matter with up, down and strange quarks
(the charm, bottom and top quarks are too massive to appear inside
neutron stars) might have a greater binding energy per baryon at zero
pressure than iron nuclei have. If true, such matter is the ultimate ground
state of matter. Normal matter is then metastable, and compressed to
sufficiently high density, would spontaneously convert to deconfined
quark matter. Unlike normal stars, 
SQM stars are self-bound, not requiring gravity to hold
them together.  It is generally assumed that
pulsars and other observed neutron stars are normal neutron stars. If
SQM stars have a bare quark surface, calculations suggest 
suggest that photon emission from SQM stars occurs
primarily in the energy range $30~{\rm keV}~<E<500~{\rm keV}$~\cite{page02b}.

\section*{How Neutron Stars are Formed}

Neutron stars are created in the aftermath of the gravitational
collapse of the core of a massive star ($>8$ M$_\odot$) at the end of
its life, which triggers a Type II supernova explosion.  Newly-born
neutron stars or proto-neutron stars are rich in leptons, mostly $e^-$
and $\nu_e$ (Fig. 1). The detailed explosion mechanism of
Type II supernovae is not understood \cite{burrows00}, but it is
probable that neutrinos play a crucial role. One of the most
remarkable aspects is that neutrinos become temporarily trapped within
the star during collapse. The typical neutrino-matter cross section is
$\sigma\approx10^{-40}$ cm$^2$, resulting in a mean free path
$\lambda\approx(\sigma n)^{-1}\approx10$ cm, where the baryon number
density is $n\simeq$ 2 to 3 $n_0$. This length is much less than the
proto-neutron star radius which exceeds 20 km. The gravitational
binding energy released in the collapse of the progenitor star's white
dwarf-like core to a neutron star is about
$3GM/5R^2\simeq3\times10^{53}$ erg ($G$ is the gravitational
constant), which is about 10\% of its total mass energy $Mc^2$.  The
kinetic energy of the expanding remnant is on the order of 1$\times
10^{51}$ to 2
$\times10^{51}$ erg, and the total energy radiated in photons is
further reduced by a factor of 100. Nearly all the energy is carried off by
neutrinos and antineutrinos of all flavors in roughly equal
proportions.

Core collapse halts when the star's interior density reaches $n_0$,
which triggers the formation of a shock wave at the core's outer
edge. The shock wave propagates only about 100 to 200 km before it
stalls, having lost energy to neutrinos and from nuclear dissociation
of the material it has plowed through [stage (I) in
Fig. 1]. Apparently, neutrinos from the core, assisted perhaps by
rotation, convection and magnetic fields, eventually resuscitate the
shock, which within seconds accelerates outwards, expelling the
massive stellar mantle. The proto-neutron star left behind rapidly
shrinks because of pressure losses from neutrino emission in its periphery
(stage II). The escape of neutrinos from the interior occurs on a
diffusion time $\tau\simeq3R^2/\lambda c\approx10$ s. The neutrinos
observed from Supernova (SN) 1987A in the Large Magellanic Cloud confirmed this
time scale and the overall energy release of $\simeq 3\times10^{53}$
ergs \cite{burrows86,hirata87,bionta87,burrows87}.

The loss of neutrinos (which forces electrons and protons to combine,
making the matter more neutron-rich) initially warms the stellar
interior.  The core temperature more than doubles (stage III),
reaching $\sim 50$ MeV ($6\times10^{11}$ K). After 10 to 20 s,
however, the steady emission of neutrinos begins to cool the
interior. Because the cross section $\sigma\propto\lambda^{-1}$ scales
as the square of the mean neutrino energy, the condition $\lambda>R$
is achieved in about 50 s. The star becomes transparent to
neutrinos (stage IV), and its cooling rate accelerates.

Neutron stars have both minimum and maximum mass limits. The maximum
mass, which is of purely general relativistic origin, is unknown, but
lies in the range of 1.44 to 3 M$_\odot$. The upper bound follows from
causality \cite{rhoades74}, that the speed of sound in dense matter is
less than the speed of light, whereas the lower bound is the largest
accurately measured pulsar mass, $1.4408\pm0.0003$ M$_\odot$, in the
binary pulsar PSR 1913+16 \cite{weisberg03}. The minimum stable
neutron star mass is about 0.1 M$_\odot$, although a more realistic
minimum stems from a neutron star's origin in a supernova. Lepton-rich
proto-neutron stars are unbound if their masses are less than about 1
M$_\odot$ \cite{haensel02}.

The proto-neutron star, in some cases, might not survive its early
evolution, collapsing instead into a black hole. This could occur in
two different ways.  First, proto-neutron stars accrete mass that has
fallen through the shock.  This accretion terminates when
the shock lifts off, but not before the star's mass has
exceeded its maximum mass. It would then collapse and its neutrino
signal would abruptly cease \cite{burrows88}. If this does not occur, a
second mode of black hole creation is possible \cite{prak01}. A
proto-neutron star's maximum mass is enhanced relative to a cold star
by its extra leptons and thermal energy. Therefore, following
accretion, the proto-neutron star could have a mass below its maximum
mass, but still greater than that of a cold star.  If so, collapse to a
black hole would occur on a diffusion time of 10 to 20 s,
longer than in the first case. Perhaps such a scenario could explain
the enigma of SN 1987A. The 10 s duration of the neutrino
signal \cite{burrows86} confirmed the birth and early survival of
a proto-neutron star, yet there is no evidence that a neutron star
exists in this supernova's remnant. The remnant's observed
luminosity is fully accounted for by radioactivity in the ejected
matter \cite{fransson02}, meaning that any contribution from magnetic
dipole radiation, expected from a rotating magnetized neutron star, is
very small. Either there is presently no neutron star, or its spin
rate or magnetic field are substantially  smaller than those
of typical pulsars.  A delayed collapse scenario could account
for these observations~\cite{prak01}.

\section*{Global Structure of Neutron Stars}

Global aspects of neutron stars, such as the mass-radius ($M-R$) relation, are
determined by the equations of hydrostatic equilibrium. For a spherical object
in general relativity (GR), these are the so-called TOV 
(Tolman-Oppenheimer-Volkov) equations \cite{tov1,tov2}:
\begin{eqnarray}
{dP\over dr} &=& -{G(m(r)+4\pi r^3P/c^2)(\rho+P/c^2)\over
r(r-2Gm(r)/c^2)} \,, \nonumber \\ 
{dm(r)\over dr} &=& 4\pi\rho r^2\,,
\end{eqnarray}
where $P$ and $\rho$ are the pressure and mass-energy density,
respectively, and $m(r)$ is the gravitational mass enclosed within a
radius $r$. Although a few exact solutions are known \cite{delgaty98},
for a realistic $P-\rho$ relation (equation of state, hereafter EOS)
these equations must be numerically solved to obtain the $M-R$
relation as shown in Fig. 2. The region in Fig. 2
bounded by the Schwarzschild condition $R\le2GM/c^2$ is excluded by
general relativity, and that bounded by $R\lsim3GM/c^2$ is excluded by
causality \cite{lattimer90}. Some normal neutron star cases, such as
GS1, contain large amounts of exotica, any of which produces a
large amount of softening and relatively small radii and maximum
masses. For small masses, SQM stars are nearly incompressible
($R\propto M^{1/3}$).

For normal neutron stars the radius is relatively insensitive to the
mass in the vicinity of 1 to 1.5 M$_\odot$ unless the maximum mass is
relatively small. A simultaneous measurement of mass and radius of an
intermediate mass star could help to discriminate among the families
of possible EOS's. Perhaps two of the most important, but unknown,
astrophysical quantities are the neutron star maximum mass and the
radius of 1.4 M$_\odot$ neutron stars.

There are large variations in predicted radii and maximum masses
(Fig. 2) because of the uncertainties in the EOS near and above
$n_0$ \cite{lattimer01}. This seems paradoxical because the properties
of matter inside laboratory nuclei are thought to be well
understood. However, an important distinction between nuclear and
neutron star matter is their relative proton fraction $x$. Nuclear
matter has nearly equal numbers of neutrons and protons
($x\simeq1/2$), but neutron star matter has only a few percent
protons.  The energy can be described with a quadratic interpolation
in the proton fraction $x$~:
\begin{equation}
E(n,x)=E(n,x=1/2)+S_v(n)(1-2x)^2\,.
\end{equation}
The symmetry energy function $S_v(n)$ is uncertain, although weak constraints 
exist from ground-state masses (binding energies) and giant dipole
resonances of laboratory nuclei.  The symmetry energy of nuclei is
divided between bulk and surface contributions, which scale with
nuclear mass number as $A$ and $A^{2/3}$, respectively, but the ranges
of $A^{1/3}$ (up to 6) and $x$ in laboratory nuclei are too small to
separate them.

A consequence of this uncertainty is that different models predict up
to a factor of 6 variation in the pressure of neutron star matter near
$n_0$, even though the pressure of symmetric matter is better known,
being nearly zero at the same density.  This pressure variation
accounts for the nearly 50\% variation in predictions of neutron star
radii \cite{lattimer01}.

A potential constraint on the EOS derives from the rotation of
neutron stars.  An absolute upper limit to the neutron star spin
frequency is the mass-shedding limit, at which the velocity of the stellar
surface equals that of an orbiting particle suspended just above the
surface.  For a rigid Newtonian sphere this frequency is the Keplerian rate
\begin{equation}
\nu_K=(2\pi)^{-1}\sqrt{GM/R^3}=1833\left(M/{\rm
M}_\odot\right)^{1/2}\left(10{\rm~km}/R\right)^{3/2} {\rm~Hz}\,.
\label{nu}
\end{equation}
However, both deformation and GR effects are important. A similar
expression, but with a coefficient of 1224 Hz and in which $M$ and $R$
refer to the mass and radius of the maximum mass, non-rotating,
configuration, describes the maximum rotation rate possible for an
EOS \cite{lattimer90,friedman86,haensel95}.  We have found that
Eq. (\ref{nu}), but with a coefficient of 1045
Hz, approximately describes the maximum rotation rate for a star of
mass $M$ (not close to the maximum mass) and non-rotating radius $R$
independently of the EOS.  The highest observed spin rate, 641
Hz from pulsar PSR B1937+21 \cite{ashworth83}, implies a radius
limit of 15.5 km for 1.4 M$_\odot$.

\section*{Internal Structure and Composition}

A neutron star has five major regions, the inner and outer cores, the
crust, the envelope and the atmosphere (Fig. 3).  The
atmosphere and envelope contain a negligible amount of mass, but the
atmosphere plays an important role in shaping the emergent photon
spectrum, and the envelope crucially influences the transport and
release of thermal energy from the star's surface. The crust,
extending about 1 to 2 km below the surface, primarily contains
nuclei. The dominant nuclei in the crust vary with density, and range
from $^{56}$Fe for matter with densities less than about $10^6$ g
cm$^{-3}$ to nuclei with $A\sim200$ but $x\sim(0.1~{\rm to}~0.2)$ near the
core-crust interface at $n\approx n_0/3$.  Such extremely neutron-rich
nuclei are not observed in the laboratory, but rare-isotope
accelerators~\cite{RIA} hope to create some of them.

Within the crust, at densities above the neutron drip
density $4\times10^{11}$ g cm$^{-3}$ where the neutron chemical
potential (the energy required to remove a neutron from the filled sea
of degenerate fermions) is zero, neutrons leak out of nuclei. At the
highest densities in the crust, more of the matter resides in the
neutron fluid than in nuclei. At the core-crust interface, nuclei are
so closely packed that they are almost touching. At somewhat lower
densities, the nuclear lattice can turn inside-out and form a
lattice of voids, which is eventually squeezed out at densities near
$n_0$ \cite{lamb78}. If so, beginning at about $0.1 n_0$, there could
be a continuous change of the dimensionality of matter from
three-dimensional (3-D) nuclei
(meatballs), to 2-D cylindrical nuclei (spaghetti), to 1-D slabs of
nuclei interlaid with planar voids (lasagna), to 2-D cylindrical voids
(ziti), to 3-D voids (ravioli, or Swiss cheese in Fig. 3)
before an eventual transition to uniform nucleonic matter
(sauce). This series of transitions is known as the nuclear
pasta.

For temperatures less than $\sim 0.1$ MeV, the neutron fluid in the
crust probably forms a $^1S_0$ superfluid \cite{baym75}.  Such a
superfluid would alter the specific heat and the neutrino emissivities
of the crust, thereby affecting how neutron stars cool.  The
superfluid would also form a reservoir of angular momentum that, being
loosely coupled to the crust, could cause pulsar glitch
phenomena \cite{anderson75}.

The core constitutes up to 99\% of the mass of the star
(Fig. 3). The outer core consists of a soup of nucleons,
electrons and muons. The neutrons could form a $^3$P$_2$ superfluid
and the protons a $^1$S$_0$ superconductor within the outer core. In
the inner core exotic particles such as strangeness-bearing hyperons
and/or Bose condensates (pions or kaons) may become abundant. It is
possible that a transition to a mixed phase of hadronic and deconfined
quark matter develops~\cite{glendenning92}, even if strange quark
matter is not the ultimate ground state of matter.  Delineating the
phase structure of dense cold quark matter \cite{alford01} has yielded
novel states of matter including color-superconducting phases
with~\cite{bedaque01} and without condensed mesons \cite{alford01}.

\section*{Neutron Star Cooling}

The interior of a proto-neutron star loses energy at a rapid rate by neutrino
emission. Within 10 to 100 years, the thermal evolution time of the crust,
heat transported by electron conduction into the interior, where it is
radiated away by neutrinos, creates an isothermal structure (stage (V) in 
Fig. 1). The star continuously emits photons, dominantly in x-rays, with an
effective temperature $T_{eff}$ that tracks the interior temperature but
that is smaller by a factor of  
$\sim 100$. The energy loss from photons is swamped by
neutrino emission from the interior until the star becomes
about $3\times10^5$ years old (stage VI).

The overall time that a neutron star will remain visible to
terrestrial observers is not yet known, but there are two
possibilities: the standard and enhanced cooling scenarios. The
dominant neutrino cooling reactions are of a general type, known as
Urca processes \cite{pethick92}, in which thermally excited particles
alternately undergo beta and inverse-beta decays. Each reaction
produces a neutrino or antineutrino, and thermal energy is
thus continuously lost.

The most efficient Urca process is the direct
Urca process involving nucleons:
\begin{equation}
n\rightarrow p+e^-+\bar\nu_e\,,\qquad p\rightarrow n+e^++\nu_e\,.
\end{equation}
This process is only permitted if energy and momentum can be
simultaneously conserved. This requires that the proton to neutron
ratio exceeds 1/8, or the proton fraction $x\ge1/9$, which is far
above the value found in neutron star matter in the vicinity of
$n_0$. In a mixture of neutrons, protons and electrons, the proton
fraction $x$ in beta equilibrium satisfies \cite{lattimer91}
\begin{equation}
x\simeq0.048~(S_v(n)/S_v(n_0))^3~(n_0/n)~(1-2x)^3\,,
\label{x}
\end{equation}
where, typically, $S_v(n_0)\simeq30$ MeV.  Because $x$ generally
increases with density, the direct Urca process might still occur
above some density threshold. However, if the direct process is not
possible, neutrino cooling must occur by the modified Urca process
\begin{equation}
n+(n,p)\rightarrow p+(n,p)+e^-+\bar\nu_e\,,\qquad p+(n,p)\rightarrow n+(n,p)
+e^++\nu_e\,,
\end{equation}
in which an additional nucleon $(n,p)$ participates in order to
conserve momentum.  The modified Urca rate is reduced by a factor of
$(T/\mu_n)^2\lsim 10^{-4}$ to $10^{-5}$ compared to the direct Urca
rate, and neutron star cooling is correspondingly slower. The standard
cooling scenario assumes that direct Urca processes cannot occur, and
predicts that neutron stars should remain observable by surface
thermal emission for up to a few million years.

The question of whether or not the direct Urca process
occurs in neutron stars is of fundamental importance. The density
dependence of the symmetry energy function $S_v$ determines the values
of $x$ and the threshold density at which the nucleonic direct Urca
process occurs (Eq. (\ref{x})). It also plays an essential role in
determining the threshold densities of other particles, such as
hyperons, pions, kaons or quarks, whose existences trigger other
direct Urca processes  \cite{pethick92}. If a star's central density
lies below the Urca threshold, enhanced cooling cannot occur. 
Again, the quantity $S_v(n)$ plays a crucial role for
neutron stars, and its inherent uncertainty means that it is presently
unknown if direct Urca processes can occur in neutron stars.

There are two additional issues affecting cooling trajectories of
neutron stars: superfluidity \cite{tsuruta98,yakovlev99} and
envelope composition \cite{chabrier97}.  Superfluidity 
quenches cooling from the direct Urca process.  However,
an additional cooling source from the formation and breaking of
nucleonic Cooper pairs increases the cooling rate from the
modified Urca process \cite{flowers76}.  Nevertheless, a clear
distinction remains between enhanced and standard cooling trajectories.

Envelope composition also plays a role in the inferred surface
temperatures.  Although it is commonly assumed that the envelope is
dominated by iron-peak nuclei, this may not be the case. Light
elements (H or He) have smaller photon opacities which enhance
surface photon emission.  Neutron stars appear warmer with
light-element envelopes for their first 100,000 years of cooling
but eventually the situation reverses \cite{page04}.

\section*{Observations and Inferred Stellar Properties}

\subsection*{Masses}

The most accurately measured neutron star masses are from timing
observations of radio binary pulsars \cite{manchester77}. These
include pulsars orbiting another neutron star, a white dwarf or a
main-sequence star.  Ordinarily, observations of pulsars in binaries
yield orbital sizes and periods from Doppler shift phenomenon, from
which the total mass of the binary can be deduced. But the compact
nature of several binary pulsars permits detection of relativistic
effects, such as Shapiro delay \cite{shapiro64} or orbit shrinkage due
to gravitational radiation reaction, which constrains the inclination
angle and permits measurement of each mass in the binary. A
sufficiently well-observed system, such as the binary pulsar PSR
1913+16 \cite{weisberg03} or the newly discovered pulsar binary PSR
J0737-3039 \cite{lyne04}, can have masses determined to impressive
accuracy. Masses can also be estimated for neutron stars that are
accreting matter from a stellar companion in so-called x-ray binaries,
but the measurements have much larger relative errors Table \ref{htable}.
%(Fig. 5).  
Neutron stars in binaries with white dwarf
companions have a broader range of masses than binary neutron stars
and the wider mass range may signify a wider range of formation
mechanisms. It has been suggested that a rather narrow set of
evolutionary circumstances conspire to form double neutron star
binaries \cite{bethe98}.  The largest apparent masses are in the
systems 4U1700-37, which might in fact contain a black hole, not a
neutron star, Vela X-1, and the pulsar J0751+1807, but all have large
uncertainties.  Raising the limit for the neutron star maximum mass
could eliminate entire EOS families, especially those in which exotica
appear and substantial softening begins around 2 to 3 $n_0$. This
could be significant, because exotica generally reduce the maximum
mass appreciably.

\subsection*{Thermal Emission}

Most known neutron stars are observed as pulsars and have photon
emissions from radio to x-ray wavelengths dominated by non-thermal
emissions. It is believed that the bulk of the non-thermal emissions
are generated in a neutron star's magnetosphere. Although such emissions
can teach us about magnetospheric phenomena, they are difficult to
utilize in constraining the star's global aspects, such as
mass, radius and temperature, that have significant bearing on a
star's interior structure, composition and evolution. About a dozen neutron
stars with high thermal emissions, and with ages up to a
million years, have been identified \cite{page04} and these stars are
expected in the standard cooling scenario to have surface temperatures
in the range of $3\times10^5$  to $10^6$ K (Fig. 4), so the
bulk of their emitted radiation should lie in the extreme ultraviolet
or x-ray regions.

The effective temperature $T_{eff,\infty}$ is defined from
\begin{equation}
F_\infty=L_\infty/4\pi d^2=\sigma_B T_{eff,\infty}^4
\left(R_\infty/d\right)^2,
\end{equation}
where $\sigma_B$ is the Stefan-Boltzmann constant, $d$ is the
distance, and $F_\infty$ and $L_\infty$ refer to the flux and
luminosity observed at Earth. These latter quantities, and
$T_{eff,\infty}$, are redshifted from the neutron star surface, where
the redshift is $z=(1-2GM/Rc^2)^{-1}-1$. For example,
$T_{eff,\infty}=T_{eff}/(1+z)$ and $F_\infty=F/(1+z)^2$. As a result,
the so-called radiation radius $R_\infty$, a quantity that can be
estimated if $F_\infty, T_{eff,\infty}$ and $d$ are known, is defined
to be $R_\infty=R(1+z)$. $R_\infty$ is a function of the
mass and radius of the neutron star, but if redshift information is
available, perhaps from spectral lines, $M$ and $R$ could be
separately determined.  Indeed, observation of spectral lines has
been reported from 1E 1207.4-5209 \cite{sanwal02} and EXO 0748-676
\cite{cottam02}, but the identifications of the lines are controversial
\cite{hailey02} with redshifts ranging from 0.12 to 0.35.

A serious hurdle in the attempt to determine $R_\infty$ and
$T_{eff,\infty}$ is the fact that neutron stars are not blackbodies
\cite{romani87,pavlov96}. The star's atmosphere rearranges the spectral
distribution of emitted radiation. Although models of neutron star
atmospheres for a variety of compositions have been constructed, these
are mostly limited to non-magnetized atmospheres. Pulsars, however,
are thought to have magnetic field strengths on the order of $10^{12}$ G or
greater \cite{manchester77}. The behavior of strongly magnetized
hydrogen is relatively simple, but models of magnetized heavy element
atmospheres are still in a state of infancy \cite{mori02}.

A useful constraint on models is provided by a few cases in which the
neutron star is sufficiently close to Earth for optical thermal
emission to be detected (distinguished by green boxes in
Fig. 4). These stars have optical fluxes several times less
than what a blackbody extrapolation from the observed x-rays into the
Rayleigh-Jeans optical domain would imply. This optical deficit is a
natural consequence of the neutron star atmosphere, and results in an
inferred $R_\infty$ greater than that deduced from a blackbody. In
most cases a heavy-element atmosphere adequately fits the global
spectral distributions from x-ray to optical energies while also
yielding neutron star radii in a plausible range. However, the
observed absence of narrow spectral features, predicted by
heavy-element atmosphere models, is puzzling
\cite{burwitz01,drake02}. The explanation could lie with broadening or
elimination of spectral features caused by intense magnetic fields or
high pressures.

Radius estimates from isolated neutron stars, while falling into a
plausible range, are also hampered by distance uncertainties. Pulsar
distances can be estimated by dispersion measures \cite{manchester77},
but these have uncertainties of 50\% or more. In a few cases, such as
Geminga \cite{caraveo96}, RX J185635-3754 \cite{kaplan02,walter02} and
PSR B0656+14 \cite{brisken03}, parallax distances have been obtained,
but errors are still large.

The recent discovery of thermal radiation from quiescent x-ray
bursters (involving neutron stars in binaries) in globular clusters is
particularly exciting. At first glance, it seems strange that neutron
stars in globular clusters, which are on the order of 10 billion years
old, could be hot enough to emit observable thermal
radiation. However, it is believed that recent episodes of mass
accretion from their companions has been a literal fountain of youth,
replenishing their reservoir of thermal energy \cite{brown98}.  The
measurements of radii from these stars might become relatively
precise, especially if the distances to the globular clusters in which
they are found can be refined. Values of $R_\infty$ in the range of 13
to 16 km have been estimated from the quiescent x-ray sources in the
globular clusters NGC 5139 and 47 Tuc \cite{rutledge02,heinke03}.

Theoretical cooling curves can be compared to observations if ages for
the thermally-emitting neutron stars can be estimated
(Fig. 4). The best-determined ages are those for which
dynamical information, such as observed space velocities coupled with
a known birthplace, is available.  Characteristic spin-down ages
estimated from pulsar periods $P$ and spin-down rates $\dot P$ using
$\tau_s = P/2\dot P$ \cite{manchester77} are less reliable.  In the
cases in which both kinds of age estimates are available, they are
generally discrepant by factors of 2 to 3.

Theoretical cooling tracks, for a variety of mass, radius and
superfluid properties, are relatively narrowly confined as long as
enhanced cooling does not occur \cite{page04}. These tracks are mostly
sensitive to envelope composition. When enhanced cooling is
considered, cooling tracks fall in a much wider range
(Fig. 4). Although most observed stars are consistent with the
standard cooling scenario, a few cases, espcially PSR J0205+6449 in
3C58 for which only upper limits to temperature and luminosity exist
\cite{slane02}, may suggest enhanced cooling.  Uncertainties in
estimated temperature and ages have precluded definitive restrictions
on EOS's or superfluid properties from being made.

\subsection*{Glitches}

Pulsars provide several sources of information concerning neutron star
properties. The fastest spinning pulsars yield constraints on neutron
star radii.  Ages and magnetic field strengths can be estimated from
$P$ and $\dot P$ measurements.  An additionally rich source of data
are pulsar glitches, the occasional disruption of the otherwise
regular pulses \cite{manchester77}.  Although the origin of glitches is
unknown, their magnitudes and stochastic behavior suggests they are
global phenomena \cite{link99}. The leading glitch model involves
angular momentum transfer in the crust from the superfluid to the
normal component \cite{anderson75}. Both are spinning, but the normal
crust is decelerated by the pulsar's magnetic dipole radiation. The
superfluid is weakly coupled with the normal matter and its rotation
rate is not diminished. But when the difference in spin rates becomes
too large, something breaks and the spin rates are brought into closer
alignment. The angular momentum observed to be transferred between
these components, in the case of the Vela pulsar, implies that at
least 1.4\% of the star's moment of inertia resides within the crust
\cite{link99}, leading to the $M-R$ limit in Fig. 2.  However,
observations of long-period ($\sim 1$ year) precesssion in isolated
pulsars appear to be inconsistent with the crustal glitch model
\cite{link03}.

\subsection*{Quasi-Periodic Oscillations}

Quasi-periodic oscillators (QPO's) are accreting neutron stars that
display quasi-periodic behavior in their x-ray emissions. Generally,
their power spectra contain a number of features the most prominent of
which are twin high frequency peaks near 1 kHz, separated by about 300
Hz.  An early interpretation of these peaks, offered in the sonic
point beat-frequency model \cite{psaltis98}, implies a
relatively large neutron star mass, $M\lsim2$ M$_\odot$
\cite{vanderklis00}.  This model holds that the higher peak frequency
is the orbital frequency of the inner edge of the accretion disk and
that the separation of the peaks is either once or twice the neutron
star's spin rate, but fails to account for the observed variations in
peak separation as a function of the lower peak frequency.  Therefore,
a variety of other models, most but not all based upon rotational
phenomenona, are under consideration \cite{vanderklis00}.  However,
none of these models seems to be wholly satisfactory in explaining the
observations \cite{vanderklis00}.

\section*{Future Prospects}

Future observations of binary pulsars and isolated neutron stars hold
the promise of effective constraints on neutron star maximum masses,
radii and internal compositions. The importance of the nuclear
symmetry energy for neutron stars and supernovae has not been
overlooked by the nuclear physics community. New accelerator
experiments, including high-resolution studies of the neutron skin
thickness (which is sensitive to the symmetry energy function $S_v$)
by parity-violating electron scattering on Pb$^{208}$, are planned
\cite{jefferson}. Anticipated studies of extremely neutron-rich nuclei
with rare-isotope accelerators \cite{RIA} will probe conditions
intermediate between laboratory nuclei and neutron star matter.
Planned intermediate energy heavy-ion experiments \cite{GSI} could
establish the in-medium properties of pions and kaons that are crucial
for delimiting the extent of Bose condensation in dense
matter. Hyper-nucleus experiments \cite{KEK} will shed light on strong
interaction couplings of strangeness-bearing hyperons likely to occur
in dense matter.

A new generation of neutrino observatories also hold great potential
for studies of proto-neutron star evolution and neutron star
structure. Neutrino observations of supernovae, validated by the
serendipitous observatons of SN 1987A which yielded about 20
neutrinos, should detect thousands of neutrinos from a galactic
supernova \cite{burrows92,jung00}. This could yield neutron star
binding energies to a few percent accuracy and provide estimates of
their masses, radii, and interior compositions, as well as details of
neutrino opacities in dense matter. Neutrino fluxes from proto-neutron
stars with and without exotica (hyperons, Bose condensates and quarks)
have been investigated in \cite{burrows86,prak01}.

Gravitational radiation is expected from asymmetric spinning compact
objects, from mergers involving neutron stars and black holes, and
from gravitational collapse supernovae \cite{thorne97}.  Depending on
the internal viscous forces in rotating neutron stars, gravitational
radiation could drive an instability in r-modes of nonradial
pulsations to grow on a time scale of tens of seconds
\cite{lindblom01}.  Mergers \cite{mergers} can be observed to great
distances.  Detectors due to begin operation over the next decade,
including LIGO (Laser Interferometer Gravitational-Wave Observatory),
VIRGO (Italian-French Laser Interferometer Collaboration), GEO600
(British-German Cooperation for Gravity Wave Experiment), and TAMA
(Japanese Interferometric Gravitational-Wave Project) could see up to
hundreds of mergers per year \cite{kalogera04}. Binary mergers can
yield important information, including the masses \cite{thorne97} and
mass-to-radius ratios of the binary's components and possibly details
of their inspiraling orbits~\cite{prakash03}.

%\newpage

\newpage
\begin{table}[t]
\begin{ruledtabular}
\caption{Neutron Star Mass Measurements ($1\sigma$ uncertainties)
\label{htable}}
\begin{tabular}{llllll}
{\bf Object} & {\bf Mass (M$_\odot$)} & {\bf Ref.} &
{\bf Object} & {\bf Mass (M$_\odot$)} & {\bf Ref.} \\ 
\hline
\multicolumn{6}{c}{\it X-Ray Binaries}  \\ 
4U1700-37$^*$ &
$2.44^{+0.27}_{-0.27}$ & \cite{clark02} & Vela X-1$^\dagger$ &
$1.86^{+0.16}_{-0.16}$ & \cite{barziv01,quaintrell03} \\

Cyg X-2 & $1.78^{+0.23}_{-0.23}$ & \cite{orosz99} &  4U1538-52 &
$0.96^{+0.19}_{-0.16}$ & \cite{vankerkwijk95} \\

SMC X-1 & $1.17^{+0.16}_{-0.16}$ & \cite{vankerkwijk95}  & LMC X-4 &
$1.47^{+0.22}_{-0.19}$ & \cite{vankerkwijk95} \\

Cen X-3 & $1.09^{+0.30}_{-0.26}$ & \cite{vankerkwijk95}  & Her X-1 &
$1.47^{+0.12}_{-0.18}$ & \cite{vankerkwijk95} \\

XTE J2123-058 & $1.53^{+0.30}_{-0.42}$ & \cite{gelino03,tomsick04} & 
2A 1822-371 & $>0.73$ & \cite{jonker03} \\

\multicolumn{6}{l}{
{~~Mean $=1.53$ M$_\odot$,}  { weighted mean $=1.48$ M$_\odot$}
}\\ 
\hline 
\multicolumn{6}{c}{\it Neutron Star -- Neutron Star Binaries}\\
1518+49 & $1.56^{+0.13}_{-0.44}$ & \cite{thorsett99} &  1518+49
companion & $1.05^{+0.45}_{-0.11}$ & \cite{thorsett99} \\

1534+12 & $1.3332^{+0.0010}_{-0.0010}$ & \cite{thorsett99}  & 1534+12
companion & $1.3452^{+0.0010}_{-0.0010}$ & \cite{thorsett99} \\

1913+16 & $1.4408^{+0.0003}_{-0.0003}$ & \cite{thorsett99}  & 1913+16
companion & $1.3873^{+0.0003}_{-0.0003}$ & \cite{thorsett99} \\

2127+11C & $1.349^{+0.040}_{-0.040}$ & \cite{thorsett99}  & 2127+11C
companion & $1.363^{+0.040}_{-0.040}$ & \cite{thorsett99} \\

J0737-3039A & $1.337^{+0.005}_{-0.005}$ & \cite{lyne04}  &
J0737-3039B & $1.250^{+0.005}_{-0.005}$ & \cite{lyne04} \\

\multicolumn{6}{l}{{~~Mean $=1.34$ M$_\odot$,}  { weighted mean $=1.41$
M$_\odot$}}\\ 
\hline 
\multicolumn{6}{c}{\it Neutron Star -- White Dwarf Binaries}\\
B2303+46 & $1.38^{+0.06}_{-0.10}$ & \cite{thorsett99} &  J1012+5307 &
 $1.68^{+0.22}_{-0.22}$ & \cite{lange01} \\

J1713+0747$^*\ddagger$ & $1.54^{+0.07}_{-0.08}$ & \cite{nice03b}  & B1802-07
& $1.26^{+0.08}_{-0.17}$ & \cite{thorsett99} \\

B1855+09$^*$ & $1.57^{+0.12}_{-0.11}$ & \cite{nice03b} &  J0621+1002
& $1.70^{+0.32}_{-0.29}$ & \cite{splaver02} \\

J0751+1807 & $2.20^{+0.20}_{-0.20}$ & \cite{nice03a,nice04}  &
J0437-4715 & $1.58^{+0.18}_{-0.18}$ & \cite{vanstraten01} \\

J1141-6545 & $1.30^{+0.02}_{-0.02}$ & \cite{bailes03} &  J1045-4509 &
$<1.48$ & \cite{thorsett99} \\

J1804-2718 & $<1.70$ & \cite{thorsett99} &  J2019+2425 & $<1.51$ &
\cite{nice01} \\
\multicolumn{6}{l}{
{~~Mean $=1.58$ M$_\odot$,}  { weighted mean $=1.34$
M$_\odot$}
}\\ 
\hline 
\multicolumn{6}{c}{Neutron Star -- Main Sequence Binary}\\
J0045-7319 & $1.58^{+0.34}_{-0.34}$ & \cite{thorsett99} & 
\end{tabular}
\end{ruledtabular}
\vspace{.5cm}
\noindent {$^*$ Could possibly be a black hole, due to lack of
pulsations.} 
\hspace*{0.25in} {$^\dagger$ Data from \cite{quaintrell03} used.} \\
\noindent {$^\ddagger$ Reflects binary period-white dwarf mass
constraint from \cite{tauris99}.}

\end{table}

\clearpage

\begin{figure}
\begin{center}
\includegraphics[width = 0.85\textwidth,angle=90]{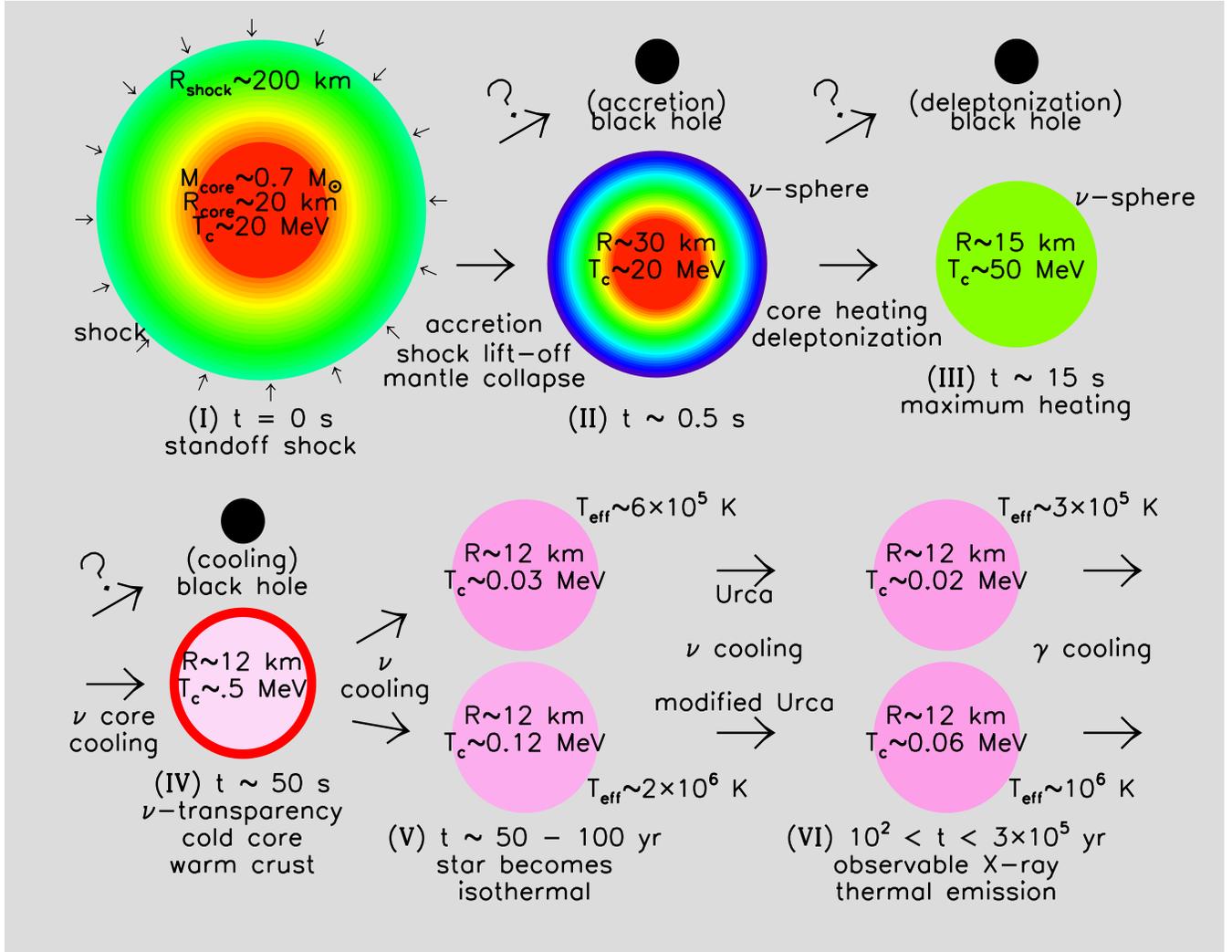}
\end{center}
\caption{The main stages of evolution of a neutron star.  Roman
numerals indicate various stages described in the text.  The radius
$R$ and central temperatures $T_c$ for the neutron star are
indicated as it evolves in time $t$.}
\label{pns}
\end{figure}

%\newpage

%
\begin{figure}
%\begin{center}
\hspace*{-0.75in}
\includegraphics[width = 0.9\textwidth,angle=90]{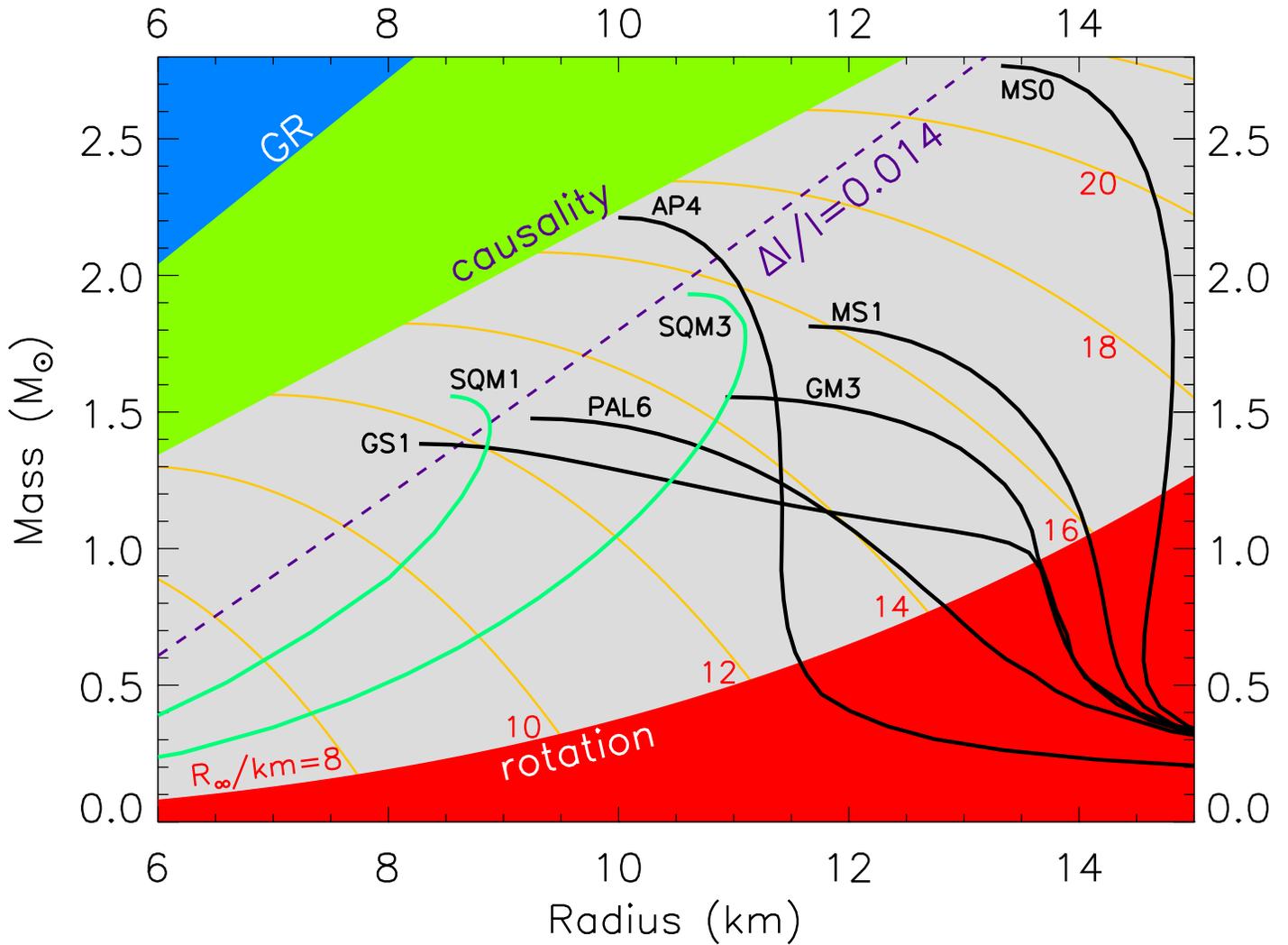}
%\end{center}
%
\caption{Mass-radius diagram for neutron stars.  Black (green) curves
are for normal matter (SQM) equations of state [for definitions of the
labels, see \cite{lattimer01}].  Regions excluded by general
relativity (GR), causality and rotation constraints are indicated.
Contours of radiation radii $R_\infty$ are given by the orange curves.
The dashed line labeled $\Delta I/I=0.014$ is a radius limit
estimated from Vela pulsar glitches \cite{lattimer01}.}
\label{mr}
\end{figure}

%\newpage

%
\begin{figure}
\begin{center}
\includegraphics[width = 0.95\textwidth,angle=0]{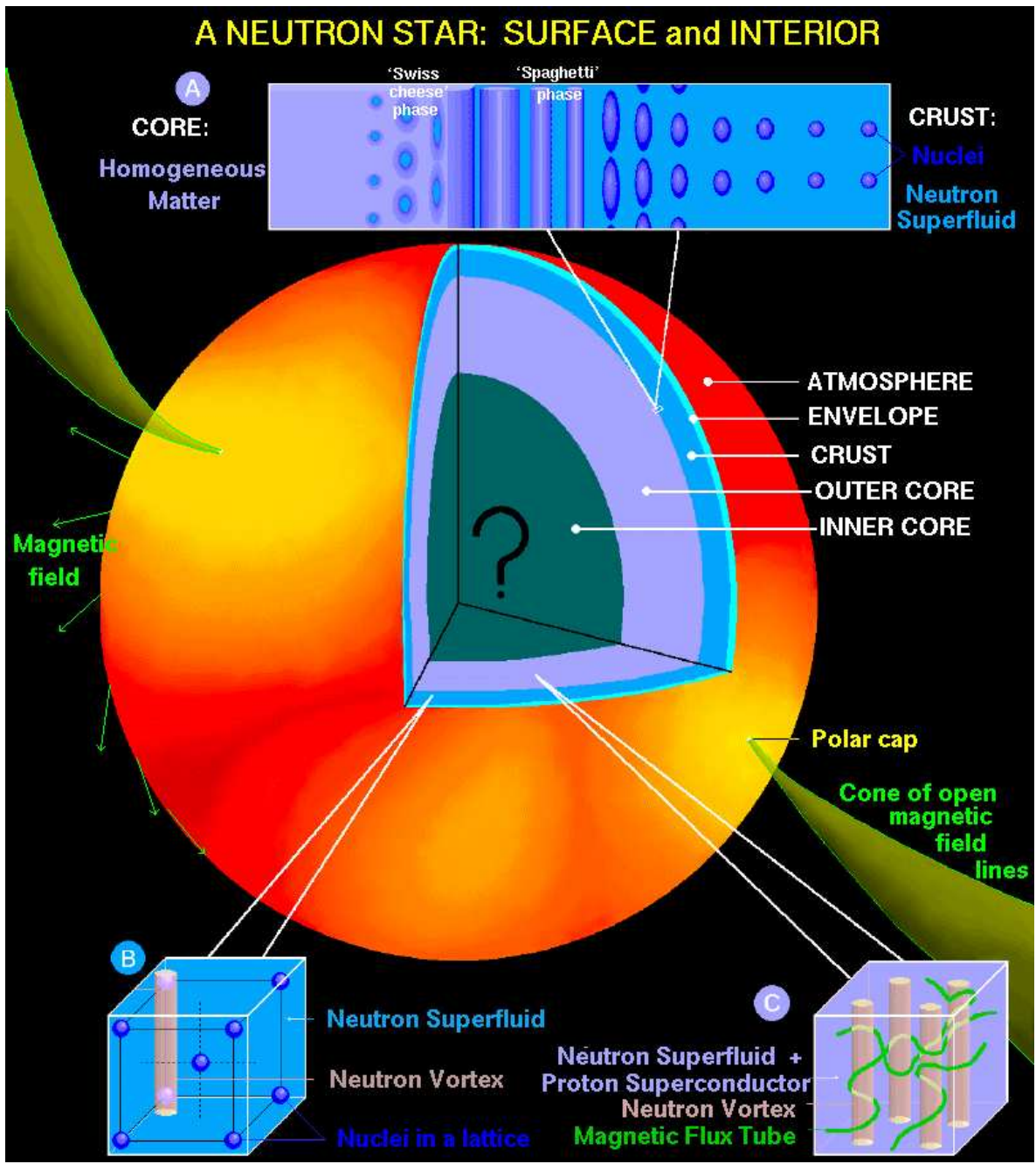}
\end{center}
\caption{The major regions and possible composition inside a normal
matter neutron star. The top bar illustrates expected
%suspected 
geometric transitions from homogenous matter at high densities in the
core to nuclei at low densities in the crust.  Superfluid aspects of
the crust and outer core are shown in insets. [Figure courtesy
D. Page.]}
\label{struct}
\end{figure}

%\newpage
%
\begin{figure}
%\begin{center}
\hspace*{-1in}
\includegraphics[width = 0.95\textwidth,angle=90]{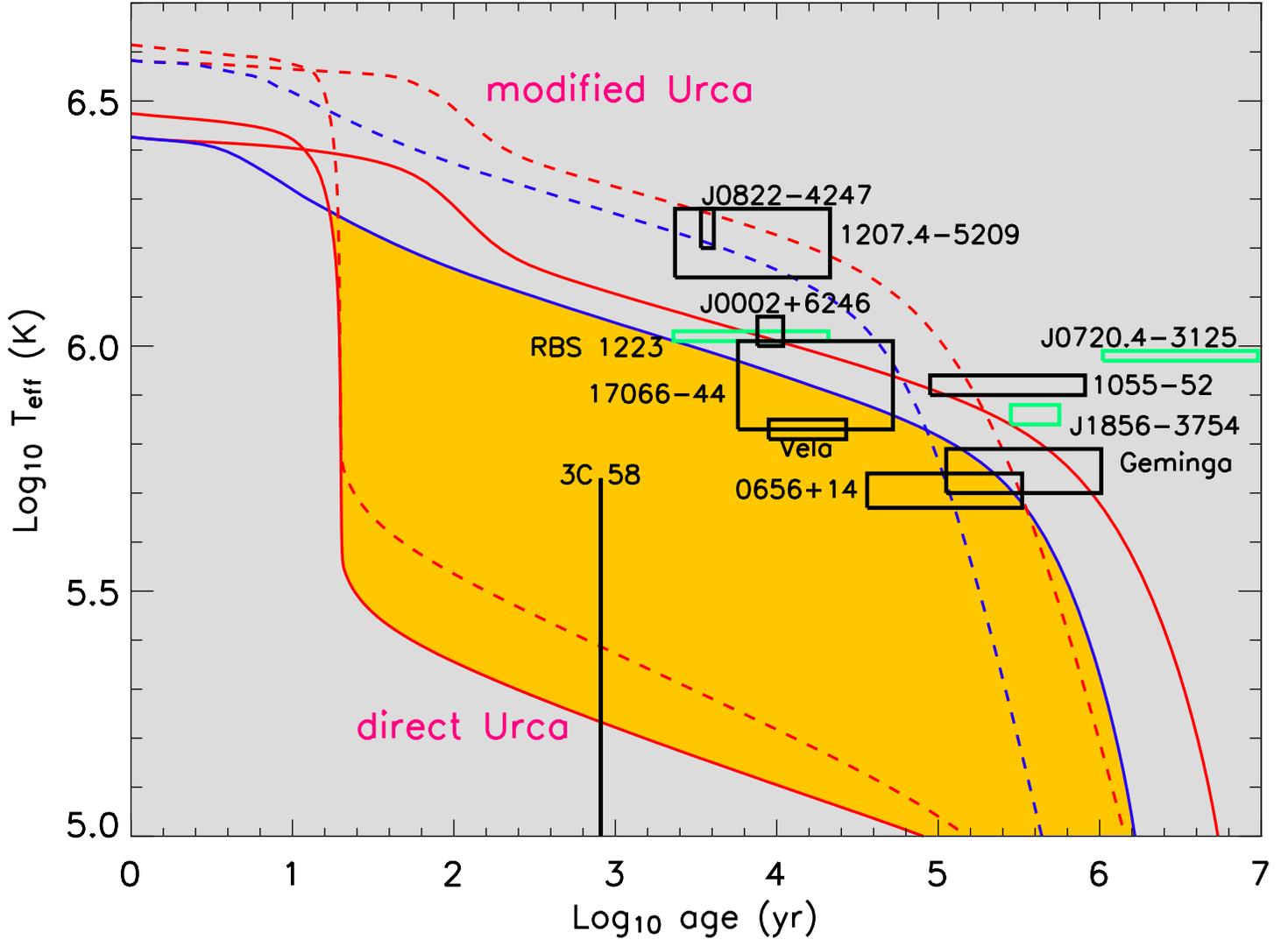}
%\end{center}
%
\caption{Observational estimates of neutron star temperatures and ages
together with theoretical cooling simulations for $M=1.4$
M$_\odot$. Models (solid and dashed curves) and data with
uncertainties (boxes) are described in \cite{page04}. The green error
boxes indicate sources from which thermal optical emissions have been
observed in addition to thermal x-rays.  Simulations with Fe (H)
envelopes are displayed by solid (dashed) curves; those including
(excluding) the effects of superfluidity are in red (blue).  The upper
four curves include cooling from modified Urca processes only, the
lower two curves allow cooling with direct Urca processes and neglect
the effects of superfluidity. Models forbidding direct Urca processes
are relatively independent of $M$ and superfluid properties. The
yellow region encompasses cooling curves for models with direct Urca
cooling including superfluidity.}
\label{cool}
\end{figure}


\begin{thebibliography}{999}
\bibitem{pethick92}The term Urca
%(unrecordable cooling agent according to Gamow and robber in Russian) 
was taken from 
%has often been linked with 
a now defunct, but once glamorous, casino of that name in Rio de Janeiro
where gamblers continuously lost money.  For a review with historical
commentary on direct and modified Urca processes, 
see \cite{pethick92n}.
\bibitem{shapiro64}Shapiro delay is the additional time required for
light to traverse the curved space near a massive object compared to
flat space.  See \cite{shapiro64n}.
\end{thebibliography}
\end{document}